\documentclass[twocolumn,twocolappendix]{aastex631}
\usepackage{pict2e}
\usepackage{amsmath}
\usepackage{times}
\usepackage{soul}
\hypersetup{linkcolor=red,citecolor=magenta,filecolor=cyan,urlcolor=magenta}

\newcommand{\ie}{i.e.,~}
\newcommand{\eg}{e.g.,~}

\newcommand{\RBM}{$R\!-\!\beta$~}

\shorttitle{Self-Consistent Electron Temperature and BH Imaging}
\shortauthors{Cruz-Osorio et al.}
\graphicspath{{./}{figures/}}

\begin{document}

\title{Supermassive black-hole imaging with a self-consistent
  electron-temperature prescription}

\correspondingauthor{Alejandro Cruz-Osorio, Claudio Meringolo}
\email{aosorio@astro.unam.mx, meringolo@itp.uni-frankfurt.de}
\author[0000-0002-3945-6342]{Alejandro Cruz-Osorio}
\affiliation{Instituto de Astronom\'{\i}a, Universidad Nacional
  Aut\'onoma de M\'exico, AP 70-264, Ciudad de M\'exico 04510, M\'exico}

\author[0000-0001-8694-3058]{Claudio Meringolo}
\affiliation{Institut f\"ur Theoretische Physik, Goethe Universit\"at,
  Max-von-Laue-Str. 1, D-60438 Frankfurt, Germany}

\author[0000-0002-1827-1656]{Christian M. Fromm}
\affiliation{Institut f\"ur Theoretische Physik und Astrophysik, Universit\"at W\"urzburg, Emil-Fischer-Str. 31, D-97074 W\"urzburg, Germany}
\affiliation{Institut f\"ur Theoretische Physik, Goethe Universit\"at,
  Max-von-Laue-Str. 1, D-60438 Frankfurt, Germany}
 \affiliation{Max-Planck-Institut f\"ur Radioastronomie, Auf dem H\"ugel 69, D-53121 Bonn, Germany}
\author[0000-0002-8131-6730]{Yosuke Mizuno}
\affiliation{Tsung-Dao Lee Institute and School of Physics and Astronomy,
  Shanghai Jiao Tong University, Shanghai, 200240, Republic of China}
\affiliation{Institut f\"ur Theoretische Physik, Goethe Universit\"at,
  Max-von-Laue-Str. 1, D-60438 Frankfurt, Germany}

\author[0000-0001-8184-2151]{Sergio Servidio}
\affiliation{Dipartimento di Fisica, Universit\`a della Calabria, I-87036
  Cosenza, Italy}

\author[0000-0002-1655-9912]{Antonios Nathanail}
\affiliation{Research Center for Astronomy and Applied Mathematics,
Academy of Athens, Athens 11527, Greece}
%\affiliation{Department of Physics, National and Kapodistrian University
%  of Athens, Panepistimiopolis, GR 15783 Zografos, Greece}
 
\author[0000-0001-9283-1191]{Ziri Younsi}
\affiliation{Mullard Space Science Laboratory, University College London,
  Holmbury St. Mary, Dorking, Surrey RH5 6NT, UK}

\author[0000-0002-1330-7103]{Luciano Rezzolla}
\affiliation{Institut f\"ur Theoretische Physik, Goethe Universit\"at,
  Max-von-Laue-Str. 1, D-60438 Frankfurt, Germany}
\affiliation{CERN, Theoretical Physics Department, 1211 Geneva 23,
  Switzerland}
\affiliation{School of Mathematics, Trinity College, Dublin 2, Ireland}

\begin{abstract}
The recent 230 GHz observations by the Event Horizon Telescope have
resolved the innermost structure of the M87 galaxy, revealing a ring-like
feature consistent with thermal synchrotron emission from a magnetized
torus surrounding a rotating supermassive black hole. Moreover, Global
Millimeter VLBI Array observations at 86 GHz have revealed a
larger-scale, edge-brightened jet with clear signatures of non-thermal
emission. The theoretical modelling of these observations involves
advanced general-relativistic magnetohydrodynamic simulations of
magnetized accretion disks around rotating black holes, together with the
associated synchrotron emission, which is normally treated with simplified
expressions for the electron temperature and assuming a purely thermal
distribution. However, an important non-thermal component is expected to
be present, making the thermal-emission model not only an approximation,
but also a source of degeneracy in the modelling. In view of this, we
here present the first application of an ab-initio approach to the
electron temperature derived from microscopic simulations of turbulent
collisionless plasmas.  The novel method, which has no tuneable
coefficients and is fully specified by the thermodynamical and
magnetic properties of the plasma, provides a better description of the
jet morphology and width at 86 GHz, as well as of the broadband spectral
emission. These findings highlight the importance of incorporating
microscopic plasma physics in black-hole imaging and emphasise the
crucial role of magnetic reconnection in electron heating and
acceleration processes.
\end{abstract}


%-------------------------------------------------------------------------
\section{Introduction}
\label{sec:intro}
%-------------------------------------------------------------------------

Over the last few years, many efforts have been dedicated to the
modelling of the emission from M87* either on the horizon scale, via its
shadow, or on parsec scales via its extended jet. The 230 GHz Event
Horizon Telescope (EHT) observations revealed a ring-like image
consistent with the predicted shadow size of a black hole (BH) in general
relativity~\citep{EHT_M87_PaperI,EHT_M87_2018}. Further observations at
86 GHz showed a ring-like structure of about $50\%$ larger than that
present at 230 GHz, and large-scale emission shows the accretion disc
connected to the edge brightened jet~\citep{Lu2023,
  Palumbo2024}. Overall, the shape and the width of the jet can be
described by taking into account its opening angle, and for M87*, the jet
base has revealed a rather wide structure~\citep{Kim2018a}. Leveraging on
17 years of Global Millimeter VLBI Array (GMVA) observations,
\citep{Kim2018a} measured the precession of the jet axis, indicating a
period of approximately 11 years in its projection on the sky. The origin
of this sideways shift remains unclear, but analytical models suggest a
precessing jet by invoking the Lense-Thirring precession of a misaligned
accretion disk~\citep{Cui2023}.

Simultaneous multi-wavelength observations have concluded that a
structured jet is necessary to explain the spectrum of
M87~\citep{Algaba2021}. A recent observational campaign in April 2018 has
shown that the flux above 350 GeV has doubled in value within a period of
$\sim$ 36 hours. At the same time, the X-ray flux was found enhanced by
about a factor of two compared to previous 2017 observations, while the
radio and millimetre core fluxes were consistent between the 2017 and
2018 campaigns. Also, evidence for a monotonically increasing jet
position angle, which corresponds to variations in the bright spot of the
EHT image, has been observed~\citep{Algaba2024}.

Among the different numerical approaches that have been considered to
model the multi-wavelength emission of M87, general-relativistic
magnetohydrodynamic (GRMHD) and general-relativistic radiative-transfer
(GRRT) calculations have been the focus of many groups worldwide
\citep{EHT_M87_PaperV, Porth2019, EHT_SgrA_PaperV, Chael2019, Cruz2022,
  Fromm2021b, Yang2024}. Overall, most of the proposed models so far need
to introduce a non-thermal electron distribution function (eDF) of some
type.

Because of the weak Coulomb coupling at collisions, the two (charged)
particle species that are normally considered in GRMHD and GRRT
simulations, namely the heavy ions and the light electrons, are expected
to have not only different temperatures, but also different energy
distributions~\citep{Tu1997, Howes2010, Dihingia2022, Zhdankin2019,
  Miceli2019, Mizuno2021}. However, GRMHD simulations can only model the
dynamics of the ``inertial'' part of the fluid, \ie of the ions, thus
leaving the knowledge about the temperature (and energy distribution) of
the electrons unknown. To tackle this problem, assumptions need to be
made about the thermodynamical properties of the electrons. In this
context, a commonly employed empirical approach is the so-called \RBM
model \citep{Moscibrodzka2016}, which has been widely used by the EHT
Collaboration to reconstruct theoretically the first images of
supermassive BH (SMBH) M87*~\citep{EHT_M87_PaperI}, and Sgr
A*~\citep{EHT_SgrA_PaperI}.

Taking into account a more realistic description of the plasma
parameters, using self-consistent kinetic models, has shown that finer
details of the image can appear, but also that the \RBM approach is
remarkably robust~\citep{Mizuno2021, Moscibrodzka2025}, albeit
rudimentary and ignoring the state of magnetisation of the plasma. Other
studies have explored the impact that the inclusion of a non-thermal
component in the eDF can have on the properties of the synchrotron
emission from the accreting plasma~\citep{Chael2017, Davelaar2019,
  Zhang2024, Tsunetoe2025}.

At the same time, Particle-In-Cell (PIC) simulations have been often used
to provide a better description of the microphysics of plasma, that is
not modeled within a fluid approach, but in terms of kinetic
theory~\citep{Comisso2018, Imbrogno2024, Imbrogno2025, Meringolo2025a}.
While computationally more expensive than GRMHD simulations, they offer a
much more realistic description of the dilute collisionless plasmas
expected near accreting SMBHs~\citep[see][who have considered this
  first]{Ball2018a}. Following this line of investigation, we here study
the impact of employing a self-consistent description for the electron
energy distribution function derived from the recent PIC simulations of
turbulent plasmas~\citep{Meringolo2023}.

The plan of the paper is as follows. An overview of the GRMHD and GRRT
simulations is presented in Sec.~\ref{sec:simulations}, where we report
the numerical parameters and the prescription used. In Sec.~\ref{sec:eDF}
we present the two approaches for the eDFs used in this work. In
Sec.~\ref{sec:results}, we report our numerical results for the
multifrequency emission of M87*, the jet morphology, the jet diameter,
and the broadband spectrum. Finally, in the last section, we discuss our
results and their implications.

%-------------------------------------------------------------------------
\section{GRMHD and GRRT simulations}
\label{sec:simulations}
%-------------------------------------------------------------------------

Our simulations of accretion onto BH and jet launching are performed
using the GRMHD code \texttt{BHAC} that solves the GRMHD equations on a
generic curved but fixed spacetime using a number of different
coordinates, in two or three dimensions, and with ability to use
block-based adaptive mesh-refinement techniques~\citep{Porth2017}. The
divergence of the magnetic field is kept at machine precision via an
accurate constrained-transport scheme~\citep{Olivares2019}. The equations
solved in \texttt{BHAC} (we use geometrised units
where $G=c=1$ and the coefficient $1/\sqrt{4\pi}$ is absorbed in the
definition of the magnetic field) are expressed as conservation laws for
the rest-mass, the conservation of mass, energy, and momentum of a
perfect fluid~\citep{Rezzolla_book:2013}, together with the Maxwell
equations for a plasma with infinite conductivity (\ie in the ideal-MHD
limit), namely
\begin{align}
  &\nabla_\mu(\rho u^\mu)=0\,, \\
  &\nabla_\mu T^{\mu \nu}=0\,, \\
  &\nabla_\mu {}^*F^{\mu \nu}=0\,.
\end{align}
Here, $\rho$ is the rest-mass density, $u^\mu$ is the four-velocity,
$T^{\mu \nu}$ is the stress-energy tensor, and ${}^*F^{\mu \nu}$ is the
dual Faraday tensor. To close the system, we employed an ideal-gas
equation of state~\cite{Rezzolla_book:2013} with an adiabatic index of
$\Gamma=4/3$ (see~\cite{Anton06, Lora2015, Porth2019, Cruz2020} for more
details).

The background spacetime corresponds to a Kerr BH expressed in
Kerr-Schild horizon-penetrating coordinates, with dimensionless spin
parameter $a_{\star}$ . The simulations were performed in spherical
coordinates $(r,\theta,\phi)$ where the grid spacing is logarithmic in
the radial direction and linear in the polar and azimuthal directions.
The dimensions of our simulation domain are set to $r\in [1.18\, M,\,
  3333\, M]$, $\theta\in [0,\, \pi]$, and $\phi \in [ 0,\, 2\pi\,]$,
where $M$ is the mass of the BH. We employ three mesh refinement
levels, resulting in an effective resolution of $(N_r,N_\theta,N_\phi )=
(512,192,192)$, respectively.

We initialise our 3D GRMHD simulations with a torus in hydrodynamic
equilibrium orbiting around the central BH~\citep[see, \eg][for
  details]{Rezzolla_book:2013}. The torus has an inner radius at $r_{\rm
  inner} =20\, M$, a maximum-pressure at the position $r_{\rm max}=40\,
M$, and constant specific angular momentum $\ell:=u_\phi/u_t=6.76$. A
weak poloidal magnetic field with a single loop is added
initially~\citep[see][for the impact of multiple loops on the
  dynamics]{Nathanail2020, Nathanail2021b} with plasma-$\beta=100$ so
that the evolution reaches the MAD state~\citep{Narayan2000}, $u_\phi$
and $u_t$ refer to the azimuthal and poloidal dual four-velocity
components, respectively. The average mass accretion rate and magnetic
Poynting flux computed in the range of time $t\in
\,[13,\!000\,M,~15,\!000\,M]$ are $\langle \dot{M}\rangle=5.2\pm 1.20$ and
$\langle \phi_{\rm BH} \rangle=33 \pm 2$, respectively~\citep[see][for
  more details]{Cruz2022, Fromm2021b}.

To generate radio images of the accretion flow around M87* and its
corresponding jet, we make use of the GRRT code
\texttt{BHOSS}~\citep{Younsi2016, Younsi2020}, which employs null
geodesics to propagate the electromagnetic radiation and solve the
radiative-transfer equation along the path of
propagation~\citep{Younsi2012}. More specifically, the radiative-transfer
equation can be written as
\begin{equation}
\dfrac{d \mathcal{I}}{d \tau_\nu} = -\mathcal{I} +
\dfrac{\eta}{\chi}\,,
\end{equation}
where $\mathcal{I} := {I_\nu}/{\nu^3}$ is the Lorentz-invariant specific
intensity, and $I_{\nu}$ is the specific intensity. The Lorentz-invariant
emissivity $\eta$, and absorptivity $\chi$, are related to the emission
$j_\nu$ and absorption coefficients $\alpha_\nu$ evaluated at frequency
$\nu$ via the definitions
\begin{align}
  &\eta := j_{0,\nu}/\nu^2\,, \\
  &\chi := \alpha_{0,\nu} \nu\,,
\label{coef}
\end{align}
where the subscript ``0'' indicates quantities measured in the local rest
frame of the plasma. The emission $j_\nu$ and absorption coefficients 
$\alpha_\nu$ are computed following \citet{Pandya2016,Marszewski2021}.

%-------------------------------------------------------------------------
\section{Electron Energy Distribution and Temperature}
 \label{sec:eDF}
%-------------------------------------------------------------------------

As mentioned in the Introduction, a commonly employed approach to
prescribe the eDF in the GRRT simulations is offered by the so-called
\RBM model~\citep{Moscibrodzka2016}, where the electrons are assumed to
have a thermal (\ie a Maxwell-J\"uttner) energy distribution and their
temperature can be deduced from that of the ions in terms of a simple
analytic function that depends on the plasma-$\beta$ parameter, \ie the
ratio of the thermal-to-magnetic pressure ($\beta := p_{\rm gas}/p_{\rm
  mag}$) and on two free parameters, \ie $R_{\rm high}$ and $R_{\rm
  low}$, which allow one to change the ``blend'' of hot electrons in the
jet region and cold electrons in the disk~\citep[see also][for a
  critical-$\beta$ model, where two additional parameters are
  introduced]{Anantua2020}. More explicitly, in the \RBM model the
electron-to-proton temperature ratio is expressed
as~\citep{Moscibrodzka2016}
\begin{equation}
  \label{eq:Tratio_RB}
  \mathcal{T}:=\frac{T_e}{T_p} = \dfrac{1+\beta^2}{R_{\rm low} + R_{\rm
      high} \beta^2}\,,
\end{equation}
where $R_{\rm low}$ and $R_{\rm high}$ are two free parameters which set
the temperature ratio in the jet (where $\beta \ll 1$) and in the disk
(where $\beta \gg 1$), respectively. In this work we set $R_{\rm low}=1$
and $R_{\rm high} = (10,160)$, since these values are routinely adopted
in the two-temperatures models \citep{Cruz2022}.

While the \RBM model is purely thermal, it is possible to ``add'' a
non-thermal part by setting a prescription for the eDF. A common approach
to do so, which we also adopt here as a comparison, is the so-called
``kappa model'' where the eDF follows a power-law distribution, \ie
\begin{equation}
  \label{eq:dne}
  \frac{d n_{\rm e}}{d \gamma_{\rm e}} = \frac{n_{\rm e}}{4 \pi}
  \gamma_{\rm e} \sqrt{\gamma_{\rm e}^2-1} \left[ 1+ \frac{\gamma_{\rm
        e}-1}{\kappa w} \right]^{-(\kappa+1)}\,,
\end{equation}
where $\gamma_{\rm e}$ is the Lorentz factor of the electrons and $N$ is
a normalization parameter. In this prescription, the weighted temperature
$w$ is defined as~\citep{Davelaar2019} 
\begin{equation}
  \label{eq:w}
  w := \frac{\kappa-3}{\kappa}\Theta_e
  +\frac{\epsilon}{2}\Big[1+\tanh(r-r_{\text{inj}})
    \Big]\frac{\kappa-3}{6\kappa}\frac{m_{\rm p}}{m_{\rm e}}\sigma\,,
\end{equation}
where $\sigma$ is the so-called ``magnetisation'' and is defined as the
ratio between the magnetic energy density and the enthalpy density, while
the dimensionless electron temperature is defined as
\begin{equation}
\Theta_{e} := \left( \frac{m_{\rm p}}{m_{\rm e}} \right)
\frac{p\,\mathcal{T}}{\rho}\,.
\end{equation}
Here $m_{\rm p}$ and $m_{\rm e}$ are the proton and electron masses and
$p$, $\rho$ refer respectively to the pressure and rest-mass density of
the ions; note that $\Theta_{e}$ should not be confused with the
electron-to-proton temperature ratio $\mathcal{T}$. Furthermore, $r_{\rm
  inj}$ in Eq.~\eqref{eq:w} is the ``injection'' radial
position~\citep[see][for details]{Fromm2021b} and $\epsilon$ is a
parameter that measures the fraction of magnetic energy that contributes
to accelerate and heat the radiating electrons and thus takes values
between zero and one.

The kappa-model aims to provide a phenomenological microscopic
description of the energy contribution from electrons accelerated by
magnetic reconnection at the base and in the cocoon -- the shocked plasma
resulting from the interaction of the jet and the ambient -- of the
jet. However, because the power-law index $\kappa$ in Eqs.~\eqref{eq:dne}
and \eqref{eq:w} is left undetermined, studies have been carried out to
express $\kappa$ in terms of the properties of the plasma. A first
important step in this direction has been taken by \citet{Ball2018a}, who
have recently carried out PIC simulations of reconnection via Harris
current sheets to obtain information of the non-thermal part of
distributions in plasma undergoing strong reconnection
processes. Collecting the data in a phenomenological expression, they
expressed the non-thermal power-law index in Eq.~(\ref{eq:dne}) as
\begin{equation}
  \kappa(\sigma,\beta) = \tilde{k}_0 + \frac{\tilde{k}_1}{\sqrt{\sigma}}
  + \tilde{k}_2 \sigma^{-0.19}\tanh \left( \tilde{k}_3 \beta
  \sigma^{0.26} \right)\,,
  \label{eq:kappa_ball}
\end{equation}
where the coefficients are ($i=0,1...,3$) $\tilde{k}_i = (2.8, 0.7, 3.7,
23.4)$. Hereafter, we will refer to this model as to the \texttt{PIC-CS}
kappa-model. As a result, using Eq.~\eqref{eq:kappa_ball} for the
non-thermal part, together with the \RBM
prescription~\eqref{eq:Tratio_RB} for the temperature ratio $\mathcal{T}$
under the assumption of a thermal distribution, it is possible to have a
reasonable first description of the electron energy distribution and
hence perform imaging of accreting SMBHs once a GRMHD simulation has been
carried out.

\begin{figure}
  \centering
  \includegraphics[width=1.0\columnwidth]{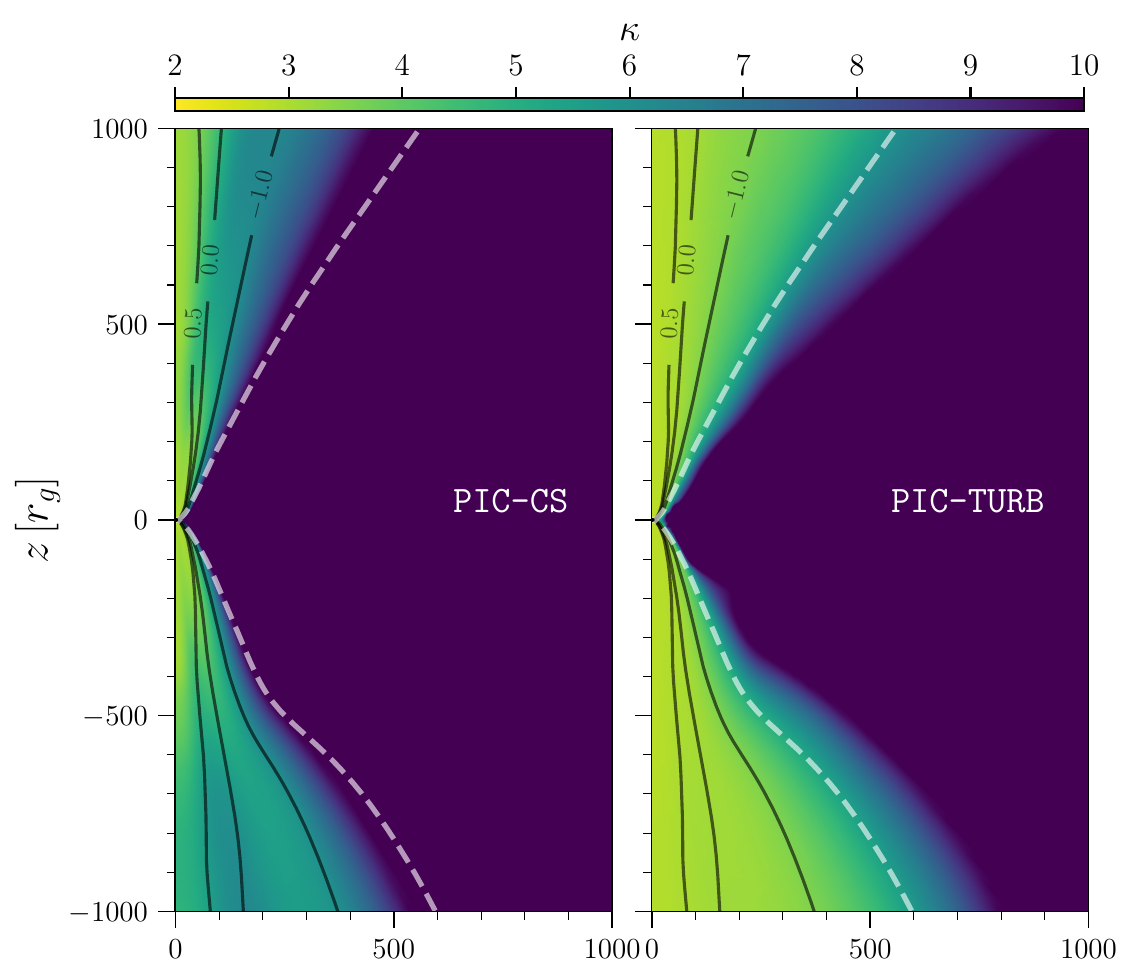}
  \includegraphics[width=1.0\columnwidth]{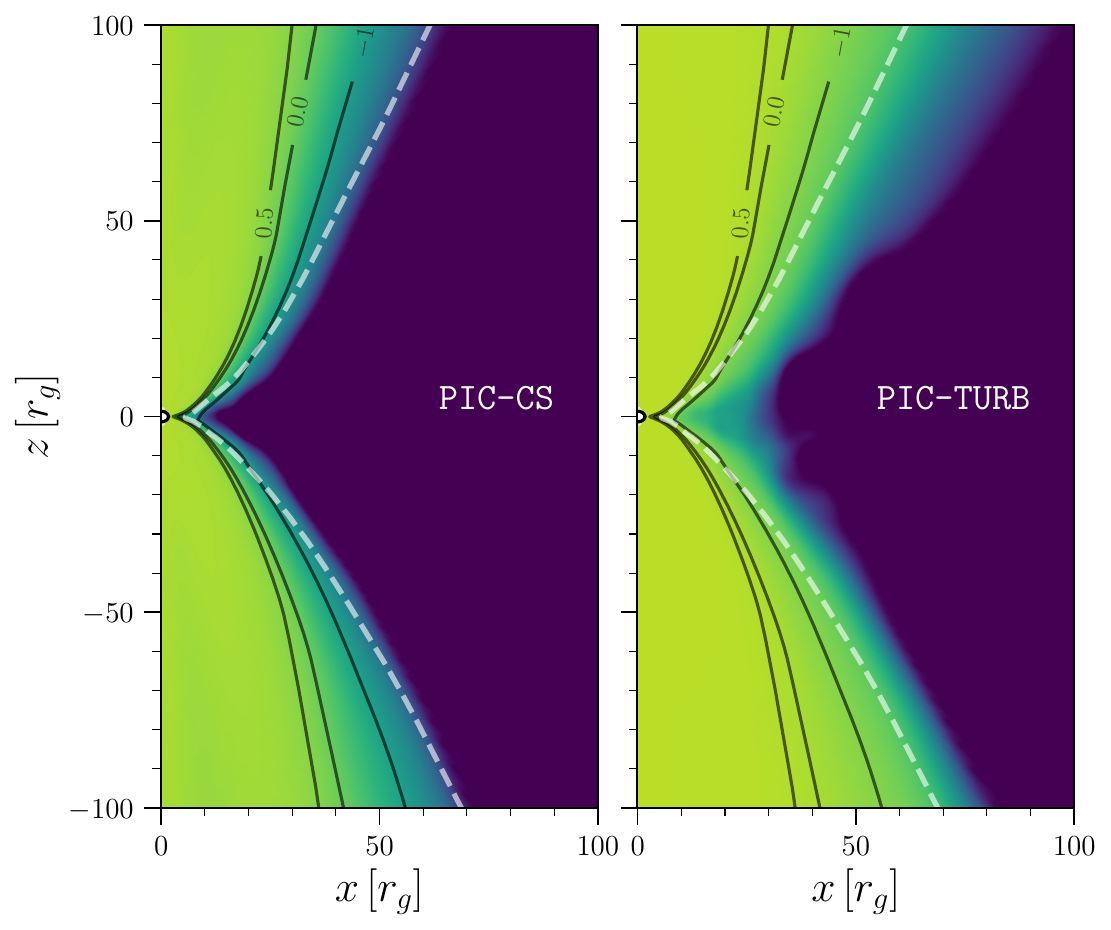}
  \caption{\textit{Top panels:} Time and azimuthally averaged
    distributions of the power-law index $\kappa$ for accretion
    simulations onto a rotating BH with dimensionless spin
    $a_{\star}=0.9375$. The left panel refers to the \texttt{PIC-CS}
    kappa-model derived from Harris current sheets
    [Eq.~\eqref{eq:kappa_ball}], while the right panel exhibits the
    distribution for the \texttt{PIC-TURB} derived from decaying plasma
    turbulence [Eq.~\eqref{eq:kappa2D_2}]. \textit{Bottom panels:} the
    same as on the top, but shown on a smaller scale near the event
    horizon. Despite the similarity in the functional dependence, the two
    models for the power-law index $\kappa$ can lead to differences,
    especially in the jet width (see also Fig.~\ref{fig:R-B_vs_PIC}). The
    black solid lines represent the separation between the jet spine and
    jet sheath and are defined by $\log_{10}\sigma=\{-1,\, 0,\, 0.5\}$;
    the dashed white line refers instead to the Bernoulli parameter ${\rm
      Be}=1.02$ and separates the bound and unbound plasma, thus marking
    external surface of the jet.}
    \label{fig:Kappa_Te}
\end{figure}

While viable, this approach does require a number of assumptions and
comes with several \textit{free parameters} that inevitably lead to a
number of possible degeneracies. To counter these limitations, and as
mentioned in Sec.~\ref{sec:intro}, a second approach is possible for the
modelling of the temperature ratio that is based on first-principles PIC
simulations. More specifically, by performing an extensive campaign of
two-dimensional, special-relativistic simulations of decaying plasma
turbulence,~\citet{Meringolo2023} were able to introduce analytic
prescriptions connecting the microphysical properties of the plasma with
the macroscopic fluid scale characteristics in the trans-relativistic
regime given by the plasma magnetisation $\sigma$ and the plasma-$\beta$
parameter. As a result, the power-law index $\kappa(\sigma,\beta)$ and
the electron-to-proton temperature ratio $\mathcal{T}(\sigma,\beta)$
were expressed in closed form as function of the macroscopic plasma
properties $\beta$ and $\sigma$, namely
\begin{eqnarray}
   \kappa (\sigma,\beta) &=& k_0 + \frac{k_1}{\sqrt{\sigma}} + k_2
   \sigma^{-6/10}\tanh \left( k_3 \beta \sigma^{1/3} \right)\,,
   \label{eq:kappa2D_2}\\
   \mathcal{T}(\sigma,\beta)  &=& t_0 + t_1 \sigma^{\tau_1} \tanh\left(t_2
   \beta\sigma^{\tau_2}\right) + t_2 \sigma^{\tau_3} \tanh\left(t_3
   \beta^{\tau_4} \sigma \right)\,, \nonumber \\ 
   \label{eq:T2D}
\end{eqnarray}
where the coefficients obtained after a multi-dimensional global fit are
$k_{0,...,3} = (2.8, 0.2, 1.6, 2.25)$, $t_{0,...,3} = (0.4,\, 0.25,\,
5.75,\, 0.037)$, and $\tau_{1,...,4}=(-0.5, 0.95, -0.3, -0.05)$,
respectively (see \cite{Meringolo2023} for details).

Note that in the limit of small magnetisation, \ie for $\beta \gg 1$ and
$\sigma \ll 1$, $\mathcal{T}(\sigma,\beta) \sim 1$ and $\kappa \gg
1$~\citep[see Fig. 4 of][]{Meringolo2023}. Under these conditions, the
eDF converges towards a purely thermal distribution. Stated differently,
the prescriptions~\eqref{eq:kappa2D_2} and \eqref{eq:T2D} naturally and
self-consistently cover all of the relevant regimes of the eDFs, from
purely thermal to highly non-thermal distribution~\citep[see also Fig. 4
  of][for a view of the range of eDFs covered by the numerical
  simulations]{Fromm2021b}. Another important point to remark is that,
since expressions~\eqref{eq:kappa2D_2} and~\eqref{eq:T2D} have been
derived from first-principles calculations and have \textit{no free
  parameters}, they represent a single prescription to describe the
complete eDF and go beyond an \RBM prescription complemented by a
\texttt{PIC-CS} model. In turn, this allows one to treat the synchrotron
emission associated with the imaging SMBH self-consistently and thus link
the microphysical properties of the plasma with the macroscopic
ones. Hereafter, we will refer to the eDF obtained using
Eqs.~\eqref{eq:kappa2D_2} and \eqref{eq:T2D} as the \texttt{PIC-TURB}
model.

\begin{figure*}[ht]
  \centering
  \includegraphics[width=0.65\textwidth]{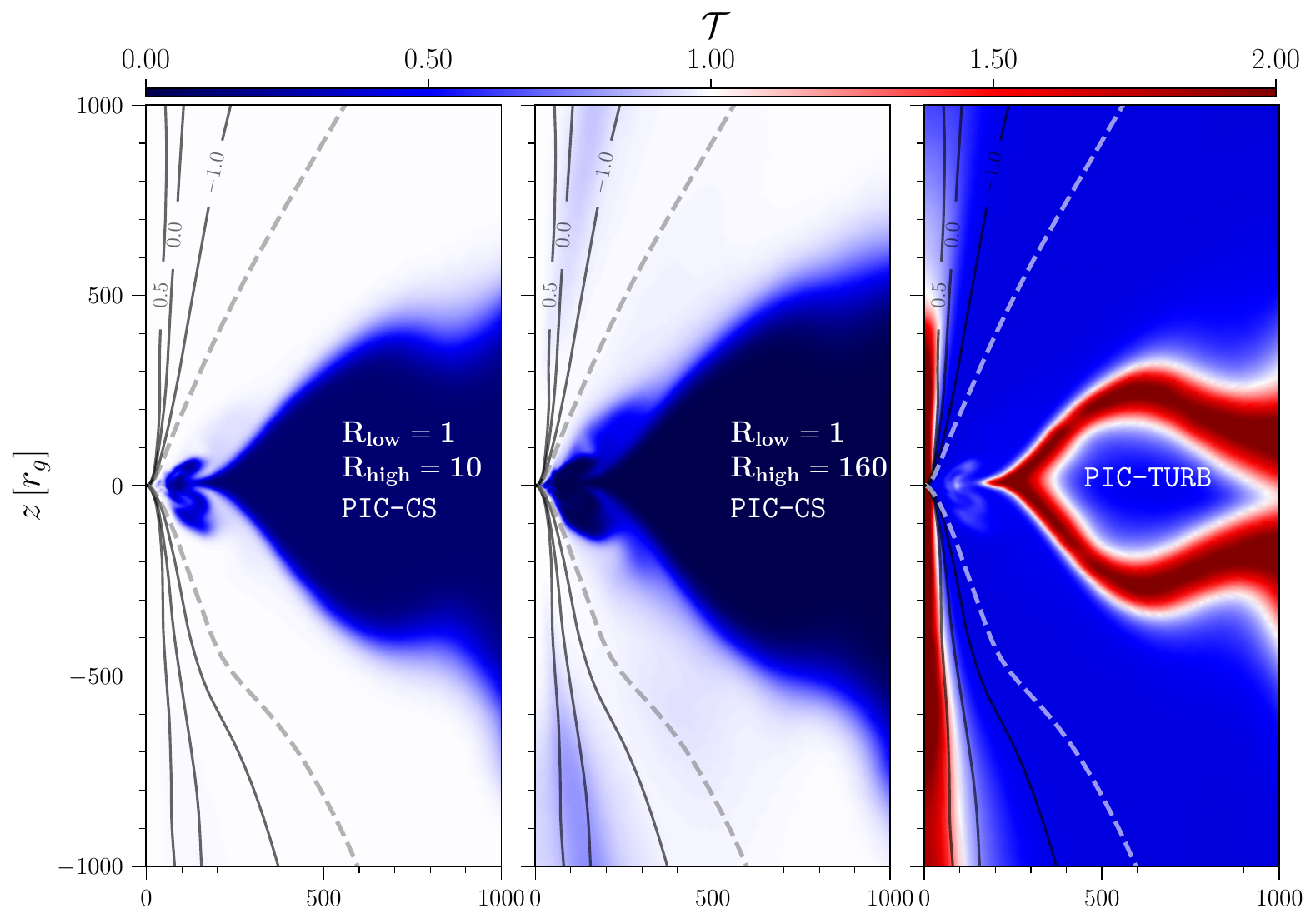}
  \includegraphics[width=0.65\textwidth]{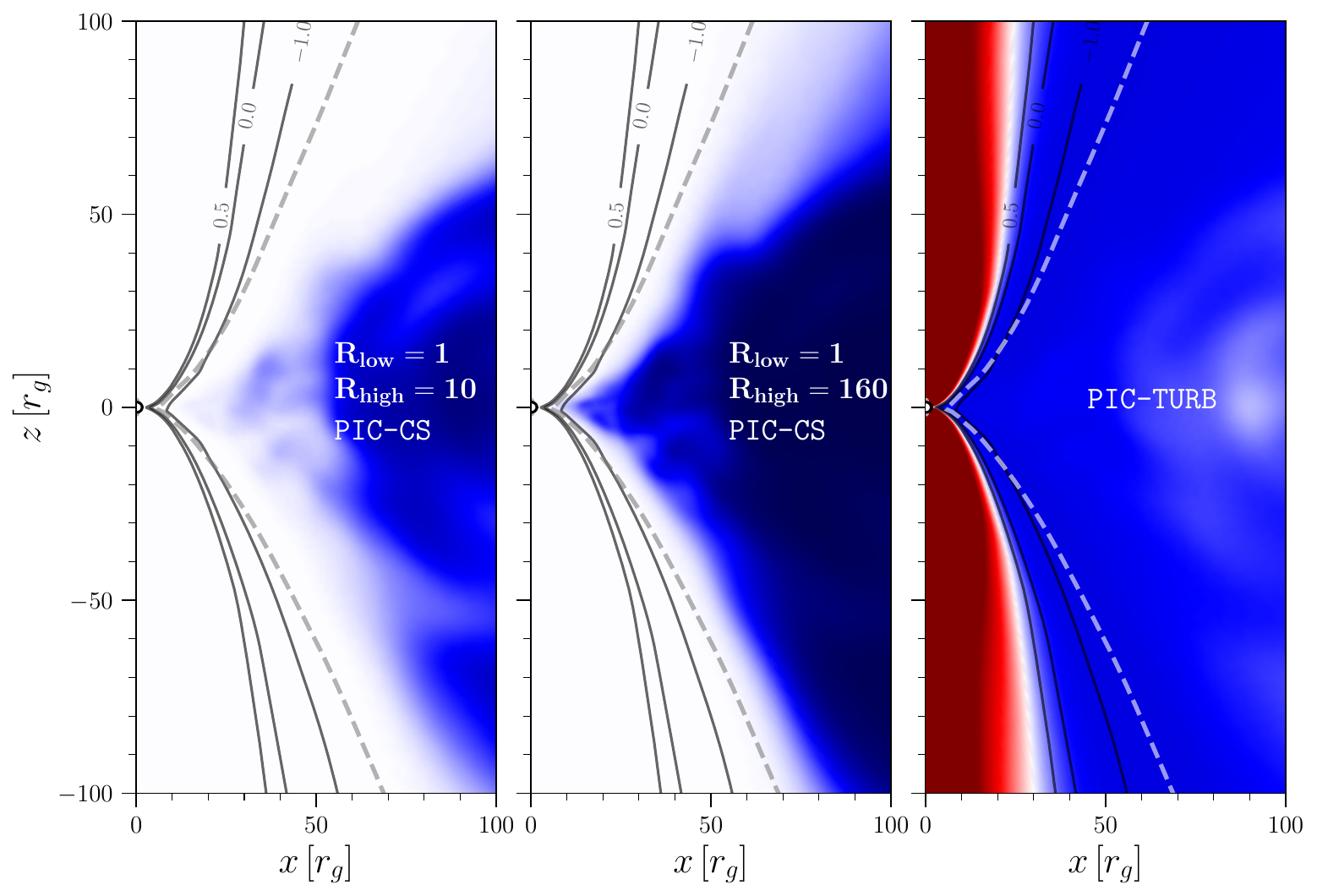}
  \caption{Comparison of the distributions in the electron temperature
    ratio, ${\cal T} := T_{e}/T_{p}$, using either the \RBM model with
    the \texttt{PIC-CS} model for the non-thermal component, or the fully
    self-consistent self-consistent prescription from turbulence
    simulations using the \texttt{PIC-TURB} model. More specifically, the
    left and middle columns report ${\cal T}$ with parameters $R_{\rm
      low}=1$ and $R_{\rm high}=\{10,\, 160\}$, while the right column
    refers to the self-consistent model. Note the considerable
    differences, especially in the core of the jet and near the torus
    surface. While the top row show the large-scale distributions, the
    bottom one offers a magnification near the event horizon. Also in
    this case, the black solid lines mark the separation between the jet
    spine and jet sheath, while the white dashed line the separation
    between bound and unbound plasma.}
  \label{fig:R-B_vs_PIC}
\end{figure*}

In the top row of Fig.~\ref{fig:Kappa_Te} we show the morphology of
plasma around a rotating BH with dimensionless spin $a_*=0.9375$. In
particular, we report the distribution of the power-law index
$\kappa=\kappa(\sigma,\beta)$ when averaged in the azimuthal direction
and over time, covering a period of $2,\!000\, M$. The white dashed line
separates the bounded and the unbounded fluid, defined by the Bernoulli
parameter ${\rm Be}:=-hu_t =1.02$. The black lines represent the contours
for the plasma magnetisation in a logarithmic scale, at the levels
$\log_{10} \sigma = \{-1,\,0,\, 0.5\}$. The latter is used as a reference
to define the jet spine, \ie for $\sigma> 3.0$. The jet sheath is then
represented by ${\rm Be} \geq1.02$ and $\sigma<3.0$. The left panels
refer to the distribution in the case of the \texttt{PIC-CS} model
of~\cite{Ball2018a} [see Eq.~\eqref{eq:kappa_ball}], while the right ones
to the distribution for the \texttt{PIC-TURB} model
of~\cite{Meringolo2023} [see Eq.~\eqref{eq:kappa2D_2}].

Note that the differences between the two models are more pronounced in
the jet region (\ie for ${\rm Be} \geq 1.02$ ) where the
\texttt{PIC-TURB} model shows smaller values of $\kappa$, representing a
broader non-thermal particles population. Far from the BH (top row), the
separation between bounded and unbounded fluid lies in $\kappa \simeq
4-6$ for the \texttt{PIC-TURB} model, while for the \texttt{PIC-CS} it is
at $\kappa \gtrsim 8$. The bottom row of Fig.~\ref{fig:Kappa_Te} offers a
magnified view of the distribution over a region closer to the BH
highlighting that the differences are more evident on these scales.

Similarly, Fig.~\ref{fig:R-B_vs_PIC} reports the distribution of the
temperature ratio $\mathcal{T}=\mathcal{T}(\sigma,\beta)$ for the same
simulation reported in Fig.~\ref{fig:Kappa_Te}, and where the bottom row
offers also in this case a magnification of the distributions in the top
one. The use of the colormap makes it rather evident that while the
temperature ratio has a rather similar distribution in the first two
panels on the left -- and referring to the \RBM model
of~\cite{Moscibrodzka2016} -- this is not the case when considering the
\texttt{PIC-TURB} model of~\cite{Meringolo2023}. Indeed, the top row of
Fig.~\ref{fig:R-B_vs_PIC} shows that in the former prescription the
temperature ratio is $\mathcal{T} \simeq 1$ in an \textit{extended} polar
region and becomes $\lesssim$ 1 only at low latitudes where the accretion
torus is present (this corresponds to $\Theta_e \sim 100$ in the jet and
$\Theta_e \sim 10^{-1}$ in the disk). This is to be contrasted with the
latter prescription, $\mathcal{T} \lesssim 2$ both in \textit{narrow}
polar region and at the surface of the torus, where higher temperatures
are naturally expected (this corresponds to $\Theta_e \sim 200$ in the
jet and $\Theta_e \sim 5$ in the disk).

These important differences become even more dramatic when considering
the zoomed-in representation in the lower row of
Fig.~\ref{fig:R-B_vs_PIC}. In this case, it is possible to realise that
the jet is considerably narrower in the self-consistent turbulence model,
resulting in a more collimated flux. Note also the presence of high
temperatures in the equatorial plane, that can be produced by a larger
number of plasmoids regions at microscopic scales, \citep[see, \eg the
  high-resolution simulations in ideal and resistive GRMHD of][where
  plasmoids are well resolved -- which enhance magnetic
  reconnection]{Ripperda2020, Ripperda2022, Nathanail2020,
  Nathanail2021b, Vos2024b, Dimitropoulos2025}. These regions enable
particles to gain energy, producing non-thermal emission.

Overall, this comparison highlights the importance of a self-consistent
description of the energy distribution in a relativistic collisionless
plasma, especially in those regions -- \ie the jet, the disk surface, and
the equatorial plane -- where it is most important for a realistic
modelling of the accretion flow. It is in fact in these regions that
physical properties of plasma directly affect the radiative properties of
the system via the emissivity and absorptivity coefficients. Our
self-consistent, kinetic prescription of the thermodynamic properties of
the electrons can thus be viewed as a way of informing the large-scale
GRMHD simulations about the small-scale influence of the plasma
turbulence.

\begin{figure*}
  \centering
  \includegraphics[width=0.99\textwidth]{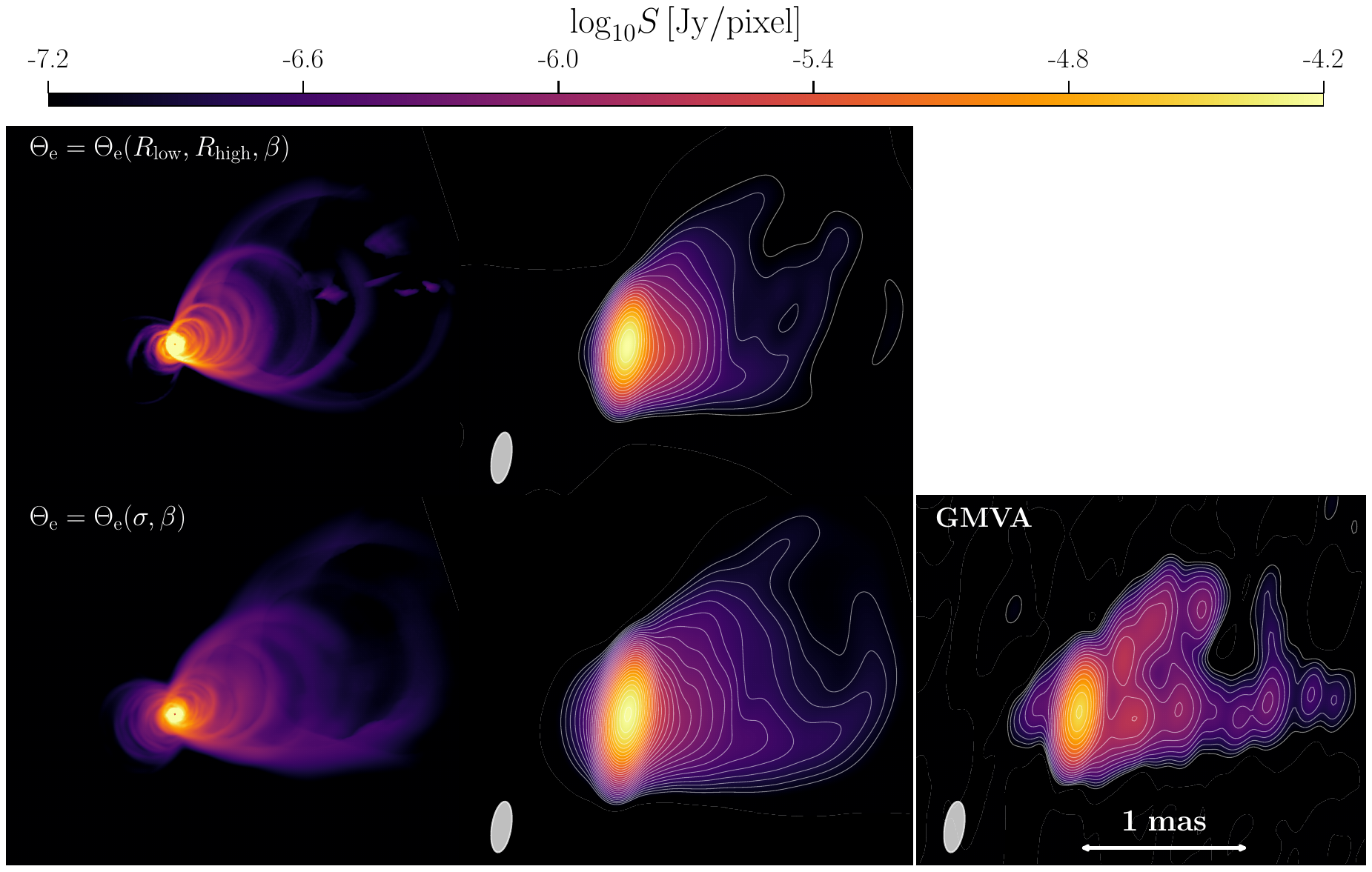}
  \caption{Best-fit models for the jet morphology of M87* as obtained
    from the most recent GMVA observations at at 86 GHz~\citep{Kim2018a}
    (bottom-right panel). The left and middle columns refer to data that
    is azimuthally and time averaged over the interval $13,\!000 -
    15,\!000,M$, for a BH with dimensionless spin $a_{\star}=
    0.9375$. The top-left panel refers to the best-fit phenomenological
    model for the electron temperature, $\Theta_e = \Theta_e(R_{\rm low},
    R_{\rm high}, \beta)$, corresponding to $R_{\rm low} = 1$ and $R_{\rm
      high} = 160$, and a $\kappa$-distribution from a current sheet with
    parameters $\epsilon = 0.5$, $\sigma_{\text{cut}} = 3.0$, $r_{\rm
      inj} = 10\,M$.  The bottom-left panel shows instead the best-fit
    model using the self-consistent electron temperature, $\Theta_e =
    \Theta_e(\sigma,\beta)$, and a $\kappa$-distribution derived from a
    turbulent scenario. In this case, with $\epsilon = 1.0$,
    $\sigma_{\text{cut}} = 3, r_{\rm inj} = 10\,M$. The middle column
    displays the GRRT synthetic image convolved with a GMVA-like beam of
    $116\,\mu{\rm as} \times 307\,\mu{\rm as}$, to mimic the
    observational resolution.}
  \label{fig:GRRT}
\end{figure*}

%-------------------------------------------------------------------------
\section{Jet morphology and multifrequency emission}
\label{sec:results}
%-------------------------------------------------------------------------

An ideal application of the new approach for the microphysics-inspired
prescription of the temperature ratio is represented by the imaging of
simulations of accreting SMBHs. In addition to the comparison with the
observational data -- both in terms of jet morphology and spectral
emission -- we also compare the new approach with the one customarily
employed by the EHT collaboration.

As a result, hereafter we will apply the results of the GRMHD simulations
discussed above to explore two different models: \textit{(i)} a
traditional one based on the thermal \RBM prescription of
\cite{Moscibrodzka2009} combined with a non-thermal eDF based on the
single current-sheet PIC simulations of \cite{Ball2018}; \textit{(ii)}
the self-consistent approach based on turbulent-plasma PIC simulations of
\cite{Meringolo2023}. While we have already mentioned this before, it is
useful to recall that approach \textit{(i)} has two tuneable and arbitrary
coefficients ($R_{\rm high}$ and $R_{\rm low}$)
[Eqs.~\eqref{eq:Tratio_RB} and~\eqref{eq:kappa_ball}]; by contrast
approach \textit{(ii)} does not have tuneable coefficients and is fully
determined by the properties of the plasma via $\sigma$ and $\beta$
[Eqs.~\eqref{eq:kappa2D_2} and~\eqref{eq:T2D}]. Because of this important
distinction, in what follows we will use the dimensionless electron
temperature $\Theta_e$ to refer to models \textit{(i)} and \textit{(ii)}
as the phenomenological $\Theta_e (R_{\rm low}, R_{\rm high}, \beta)$ and
the self-consistent $\Theta_e (\sigma, \beta)$, respectively.

%=========================================================================
\subsection{Large-Scale Jet Morphology at 86 GHz}
%=========================================================================

The focus of our attention here is therefore to reproduce with both
approaches the total-flux distribution of M87* at 230 GHz of $\simeq
1.0\,{\rm Jy}$ from the observation~\citep{Akiyama2015, Doeleman2012}. To
this scope, we used a bisection method to find the normalized
mass-accretion rate, by assuming the BH mass of $6.5 \times 10^9\,
M_\odot$ and a distance of $16.8\,{\rm Mpc}$. During the GRRT
calculations we computed the jet emission in a field of view of $4\, {\rm
  mas}\ \sim 10^3\,M$ with resolution of $800\times800$ pixels, and an
inclination angle of the observation of
$163^{\circ}$~\citep{EHT_M87_PaperV}. Furthermore, to reproduce the
morphology of the M87* jet , we have explored a set of models varying key
physical parameters. Specifically, we considered five BH spins,
$a_{\star} = [\pm 15/16, \pm 1/2, 0]$, four electron injection radii,
$r_{\rm inj} = [10, 100, 200, 400]\,M$, and five values of the magnetic
dissipation efficiency into electron heating, $\epsilon=[0,
  0.25,0.5,0.75,1.0]$.

\begin{figure}[t]
  \centering \includegraphics[width=0.45\textwidth]{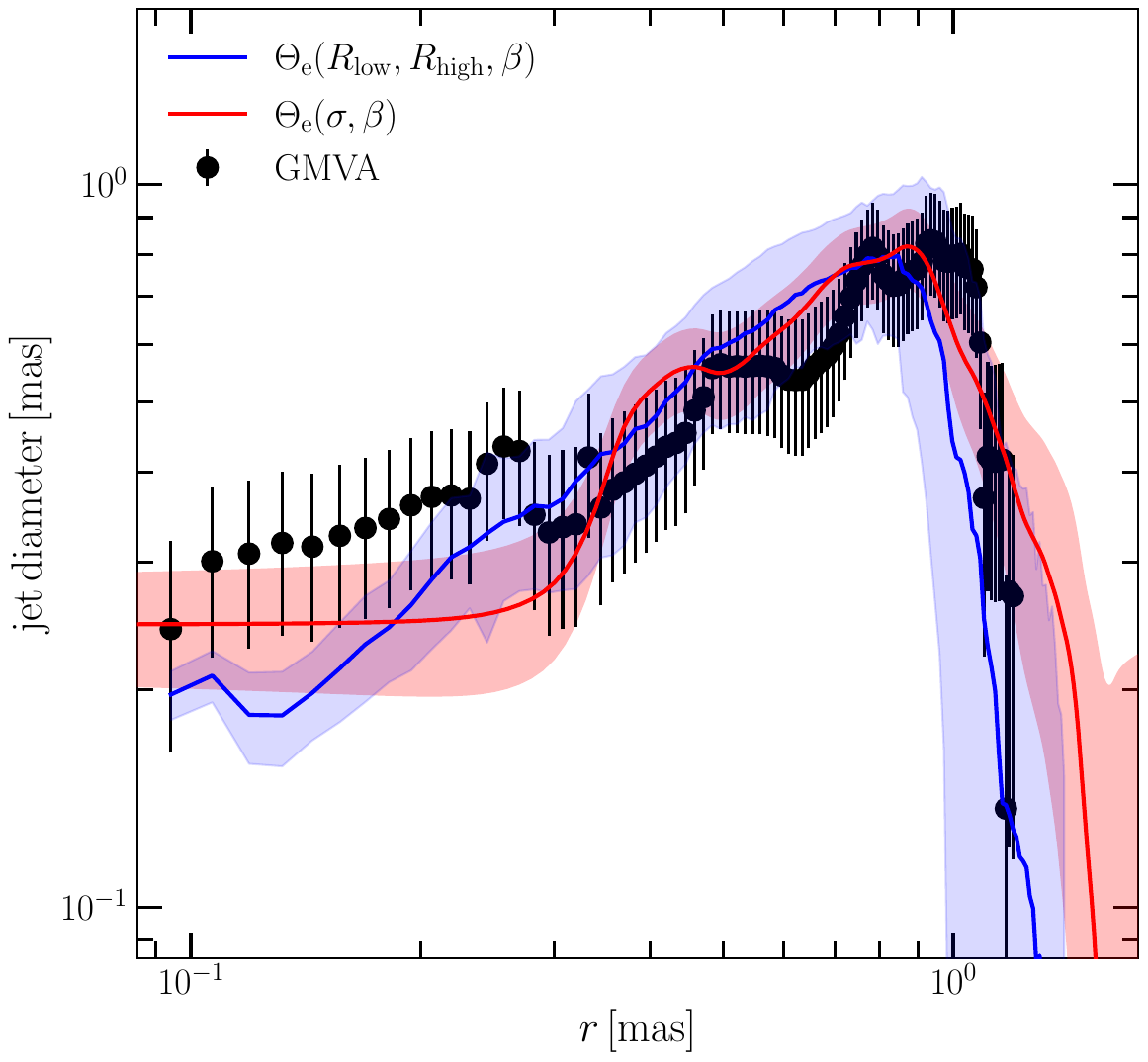}
 \caption{Jet-diameter comparison between observations and theoretical
   models. We report with solid lines the jet width measures from the
   convolved GRRT images covering a period of $2,\!000 M$, red line for
   the self-consistent $\Theta_e= \Theta_e(\sigma,\beta)$, and blue line
   for the phenomenological $\Theta_e = \Theta_e(R_{\rm low}, R_{\rm
     high}, \beta)$. We also show as shaded regions the variations within
   the standard deviation and with black circles the jet width as
   computed from the GMVA observation and the corresponding
   uncertainties. The uncertainties are obtained by assuming an
   uncertainty of 1/4 of the beam size at the $r = 0$ and a linear
   increase until 1/2 of the beam size is reached at $r = 2 \, \mu{\rm
     as}$.}
 \label{fig:jet}
\end{figure}

In Figure~\ref{fig:GRRT}, we present our best-fit models for the jet
morphology of M87* as obtained from the most recent GMVA observations at
at 86 GHz~\citep{Kim2018a} (bottom-right panel). The synthetic images on
the left and middle columns were produced using GRMHD and GRRT
simulations over the time interval $13,\!000 - 15,\!000\,M$,
corresponding to approximately two years of observation. We constrain the
dynamical range of the flux-density to span three orders of magnitude
from the maximum value, \ie $\log_{10} S \in [-7.2, -4.2]\,{\rm
  Jy/pixel}$, thus matching the GMVA observations.

In the top-left panel we present the best-fit phenomenological model for
the electron temperature, $\Theta_e = \Theta_e(R_{\rm low}, R_{\rm high},
\beta)$, corresponding to $R_{\rm low} = 1$ and $R_{\rm high} = 160$, and
a $\kappa$-distribution from a current sheet with parameters $\epsilon =
0.5$, $\sigma_{\text{cut}} = 3.0$, $r_{\rm inj} = 10\,M$. This imaging of
the total-flux emission is to be contrasted with the one presented in the
bottom-left panel of Fig.~\ref{fig:GRRT}, which shows instead the
best-fit model using the self-consistent electron temperature, $\Theta_e
= \Theta_e(\sigma,\beta)$, and a $\kappa$-distribution derived from a
turbulent scenario. In this case, with $\epsilon = 1.0$,
$\sigma_{\text{cut}} = 3, r_{\rm inj} = 10\,M$. In both approaches, the
best match is found for a BH spin of $a_{\star}= 0.9375$.

The first column of Fig.~\ref{fig:GRRT} displays instead the GRRT
synthetic image, while the second column shows the image convolved with a
GMVA-like beam of $116\,\mu{\rm as} \times 307\,\mu{\rm as}$, to mimic
the observational resolution. When comparing the old and new approach it
emerges clearly that self-consistent approach results in a more extended
and broader counter-jet, as well as in a brighter and more elongated
forward jet. Both of these properties better matches the observed
structure. Furthermore, an enhanced limb-brightening is also evident,
consistent with features seen in the data. The very good match with the
data from the self-consistent microscopic model without any intensive
tuning of the coefficients $R_{\rm low}$ and $R_{\rm high}$ is both
comforting and rewarding. It highlights that a non-parametric, physically
motivated description of the jet physics is possible and that it leads to
a robust reproduction of the observational data.

%=========================================================================
\subsection{Jet Width at 86 GHz}
%=========================================================================

A more quantitative comparison between the observed and simulated jet
structures can be made by measuring the jet diameter along the direction
of propagation. To this end, we extract various sections along the
direction that is orthogonal to that of the jet propagation and fit the
resulting flux-density profiles with Gaussian functions~\citep[see][for
  more details]{Cruz2022,Fromm2021b}. In this way, it is possible to
estimate the ``jet width'' along the propagation direction and we present
in Fig.~\ref{fig:jet} its progression as a function of the distance from
the BH for the best-fit models shown in Figure~\ref{fig:GRRT}. More
specifically, the blue line corresponds to the phenomenological $\Theta_e
= \Theta_e(R_{\rm low}, R_{\rm high}, \beta)$ model, the red line to the
self-consistent $\Theta_e = \Theta_e(\sigma,\beta)$ model, and the black
circles represent the observational GMVA data with the corresponding
error bars that are obtained by assuming an uncertainty of $1/4$ of the
beam size at the $r=0$ and a linear increase until $1/2$ of the beam size
is reached at $r=2\,{\rm mas}$.

Note that at small radial distances (\ie $r \lesssim 0.3\, {\rm mas}$),
both models underestimate the jet width compared to the observations,
although the new self-consistent model provides a more accurate
reproduction of the data. At larger distances (\ie $0.3 \lesssim r
\lesssim 1.0$ mas), both models reproduce the observed jet widths
reasonably well, with reduced chi-square values that are $\chi^2 = 1.39$
for the $\Theta_e = \Theta_e(\sigma,\beta)$ model and $\chi^2 = 1.46$ for
the $\Theta_e(R_{\rm low}, R_{\rm high}, \beta)$ model.

%=========================================================================
\subsection{Broadband Spectral Energy Distribution}
%=========================================================================

\begin{figure}[t]
  \centering \includegraphics[width=0.48\textwidth]{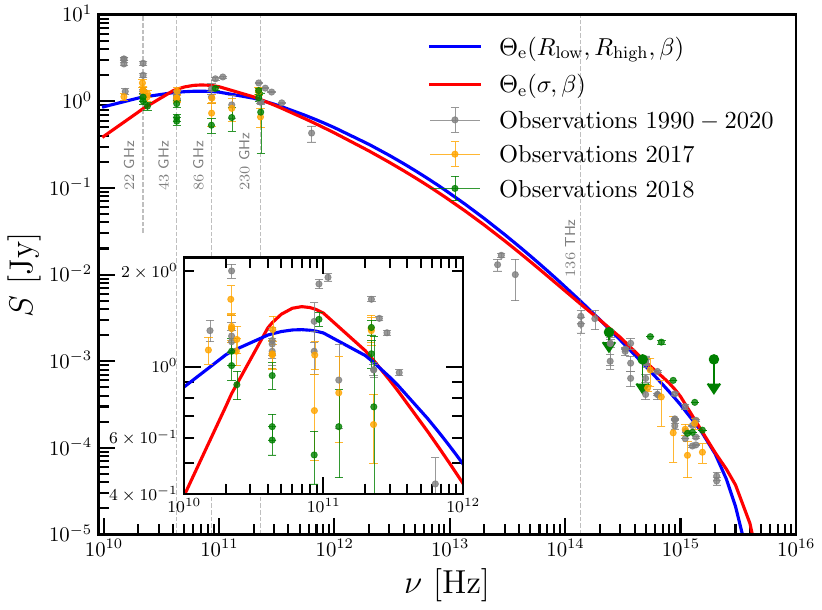}
  \caption{Broadband spectrum of the flux-density of M87*. Solid lines
    show the average spectra from simulations with different spins and
    numerical parameters, as computed over the time window between
    $13,\!000\, M$ and $15,\!000\,M$ for both non-thermal models and for
    an inclination $i = 160 ^{\circ}$. Note that the gray vertical lines
    refer to the most representative frequencies. For each observational
    data, the uncertainties indicate the variability during the
    observations. The inset shows a magnification of the low-frequency
    region.}
  \label{fig:SED}
\end{figure}

For completeness, in Fig.~\ref{fig:SED} we present the Spectral Energy
Distribution (SED) corresponding to the best-fit models. Our analysis
focuses on the first hump of the M87* spectrum, spanning the radio,
optical, near-infrared, and soft X-ray bands -- \ie from
$10^{10}\,\mathrm{Hz}$ to $10^{16}\,\mathrm{Hz}$ -- where synchrotron
emission dominates. The figure includes historical observations spanning
over 30 years, shown as grey dots with error bars indicating observed
variability. Additionally, we include simultaneous multi-wavelength
observations obtained in 2017 (orange dots) and 2018 (green circles), as
reported by~\citet{Prieto2016},~\citet{Algaba2021},
and~\citet{Algaba2024}.

\begin{figure*}
 \centering \includegraphics[width=0.8\textwidth]{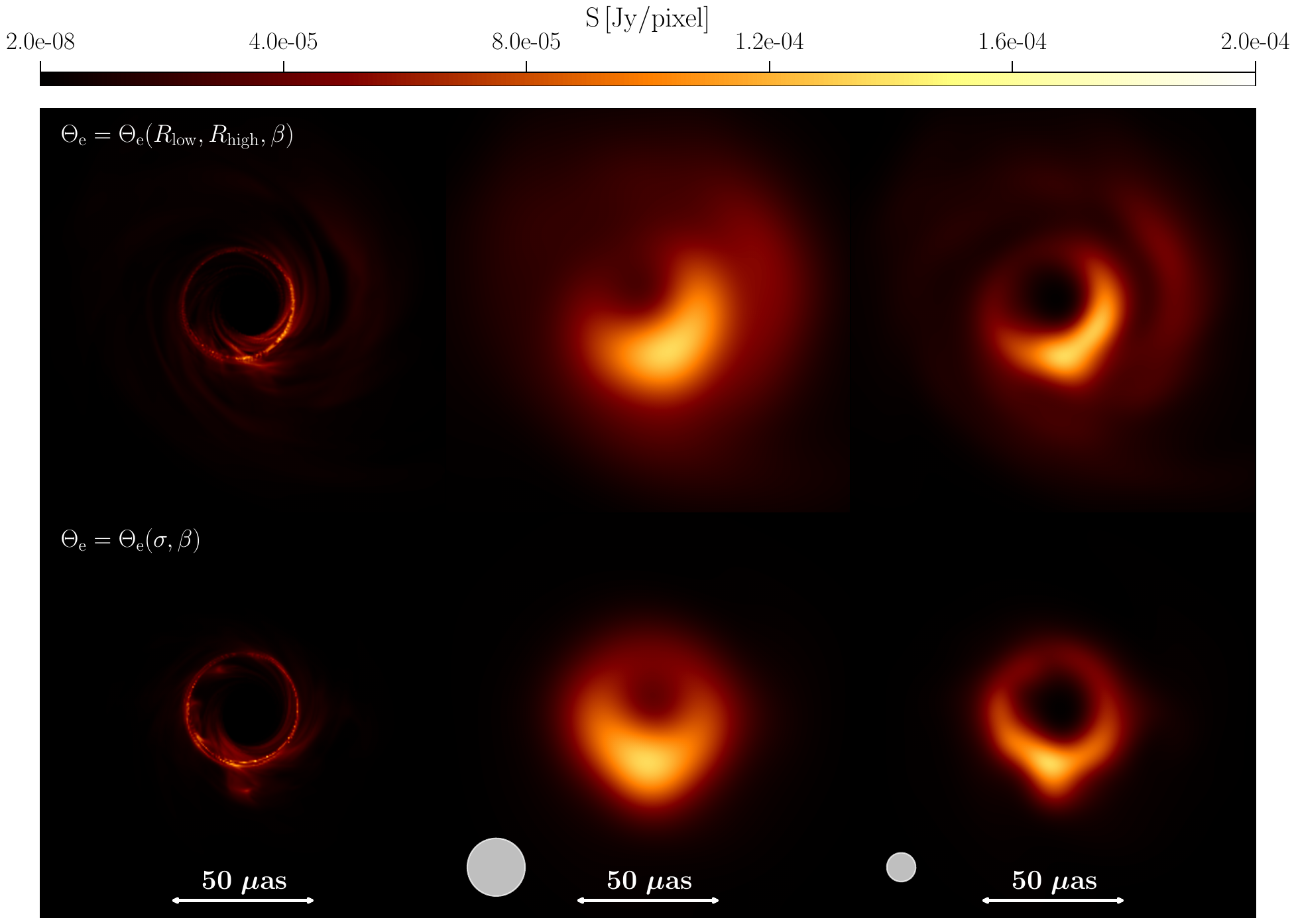}
 \caption{GRRT (left panels) and convolved images (middle and right
   panels) for our best-fit models, using either the phenomenological
   prescription $\Theta_e=\Theta_e(R_{\rm low}, R_{\rm high},\beta)$
   (first row) or the self-consistent prescription
   $\Theta_e=\Theta_e(\sigma,\beta)$ (bottom row). The panels in the
   central column correspond to synthetic images at 230 GHz convolved
   with a beam with radius of $20\, \mu{\rm as}$, mimicking the
   observational resolution for M87*, while in the right column, are
   reported the images at 345 GHz, with a smaller beam of $10\, \mu{\rm
     as}$. }
 \label{fig:230GHz}
\end{figure*}

When comparing the models with the data it emerges that both the
phenomenological (blue line) and the self-consistent (red line) models
show a very good agreement with the observed SED. The primary difference
between the two lies in the self-absorption region for frequencies below
$22\,\mathrm{GHz}$, where the self-consistent model exhibits a steeper
profile. However, in the radio, optical, near-infrared, and soft X-ray
bands, both models align well with the observations. Furthermore, as
reported in the inset, both models successfully reproduce the most recent
observational data.

%-------------------------------------------------------------------------
\subsection{Horizon-scale Images at 230 and 345 GHz}
%-------------------------------------------------------------------------

As a concluding analysis, we present in Fig.~\ref{fig:230GHz} a direct
comparison between images at both 230 and 345 GHz and therefore on the
much smaller scales of the event horizon. In order to reproduce the EHT
observations, we mimic the typical angular resolutions of VLBI
measurements by convolving our GRRT images with circular Gaussian beams
of different diameters. As for Fig.~\ref{fig:GRRT}, we report in the
first row the results obtained using the phenomenological
electron-temperature prescription $\Theta_e=\Theta_e(R_{\rm low}, R_{\rm
  high},\beta)$ and in the bottom row the corresponding results for our
self-consistent $\Theta_e= \Theta_e(\sigma,\beta)$ model. We also note
that, to maintain consistency between the two models, we consider again
the same simulation snapshot at time $t=14,\!600\,M$.

The left column reports the raw GRRT images, while the central and right
columns show the convolved images for the two different observing
frequencies at an inclination of $i = 160 ^{\circ}$. The images at 230
GHz (middle column) are convolved with a circular beam of radius
$20~\mu{\rm as}$, consistent with the typical observational resolution of
the EHT for M87*~\citep{EHT_M87_PaperI, Akiyama2019_L4}. By contrast, the
345 GHz images (right column) are convolved with a smaller $10~\mu{\rm
  as}$ beam, representative of the expected resolution at higher
frequencies~\citep{Raymond2024}.

The low inclination used in Fig.~\ref{fig:230GHz} corresponds to an
almost face-on view of the BH, with the region surrounding the photon
ring including the (projected) jet region and in the corona of the
accretion torus; both regions are characterised by $\beta \ll 1$. Under
these physical conditions, the two prescriptions described by
Eqs.~\eqref{eq:Tratio_RB} and \eqref{eq:T2D} result in slightly different
electron temperatures in different regions. More specifically,
$\Theta_e(R_{\rm low}, R_{\rm high}, \beta) \gtrsim
\Theta_e(\sigma,\beta)$ in the jet sheath and in the disk corona, where
the source is more diffused due to the enhanced electron temperature of
the \RBM model, while $\Theta_e(R_{\rm low}, R_{\rm high}, \beta)
\lesssim \Theta_e(\sigma,\beta)$ near the polar axis, where the source is
more compact (see bottom row of Fig.~\ref{fig:R-B_vs_PIC}). As a result,
the flux-density computed with the self-consistent $\Theta_e=
\Theta_e(\sigma,\beta)$ model is more compact in the jet region and has a
smaller diffused emission in the disk region. Consequently, the total
flux in the jet region is less pronounced at both frequencies when
applying our self-consistent prescription $\Theta_e(\sigma,\beta)$.

These differences are even more evident for the convolved images (middle
and right panels).  More specifically, the images convolved with a
$20~\mu{\rm as}$ beam generally appear more diffused, showing a broader
emission region, a weaker central brightness depression, and smoother
flux distribution. In contrast, the smaller beam used at 345 GHz shows
finer substructure and improves the visibility of the photon
ring.

%-------------------------------------------------------------------------
\section{Conclusions}
\label{sec:conclusions}
%-------------------------------------------------------------------------

As the quality of the images of accreting SMBHs is rapidly increasing and
will continue to do so in the future~\citep[see, \eg][]{Uniyal2025}, it
is of extreme importance for the theoretical modelling of these objects
to be able to distinguish those aspects of the images that reflect
gravitational properties of the underlying spacetimes~\citep[see,
  \eg][]{Gralla2019, Kocherlakota2021b} from the astrophysical ones,
especially near the BH and in the jet sheath at horizon
scales~\citep[see, \eg][]{Kocherlakota2022, Kocherlakota2023, Hada2024,
  Moriyama2025}. To reach this goal, it is imperative that the
astrophysical modelling is treated in the most realistic manner
possible. A crucial role in this context is played by the energy
distribution of the electrons, as this enters in the calculation of the
synchrotron emission that by far dominates the electromagnetic signal
that we receive from accreting SMBHs.

However, standard, single-fluid GRMHD simulations~\citep{Porth2019} can
only track the thermodynamics of the inertial component of the plasma,
\ie the ions, leaving the specification of the energy distribution
undetermined. To address this problem, it is customary to assume an energy
distribution that is ``thermal'' and complemented at high energies by a
``non-thermal'' component. While this is an important step forward, it
still leaves a number of tuneable parameters which inevitably lead to a
potential degeneracy in the results. 

As a result, we have here investigated the impact that a first-principle
description of the energy distribution of the electrons in the accreting
plasma has on the resulting imaging and spectral energy distribution. Our
model for describing the properties of the electron temperature and
non-thermal energy distribution is derived from PIC simulations of
turbulent and collisionless plasmas, which are believed to model
realistically the physical conditions of the plasma in the vicinity of
the event horizon. Because this description does not have free parameters
but is fully determined by the physical properties of the plasma, namely,
its magnetisation $\sigma$ and the pressure ratio $\beta$, it is both
more fundamental and more ``rigid'' than alternative approaches where
tuneable parameters can be employed.

To test how well this ``rigidity'' performs when the theoretical
modelling is compared with the astronomical observations, we have
explored two different emission models: \textit{(i)} a traditional one
based on the thermal \RBM prescription of \cite{Moscibrodzka2009}
combined with a non-thermal eDF based on the single current-sheet PIC
simulations of \cite{Ball2018} (\ie the phenomenological $\Theta_e =
\Theta_e(R_{\rm low}, R_{\rm high}, \beta)$ model); \textit{(ii)} the
self-consistent approach based on turbulent-plasma PIC simulations of
\cite{Meringolo2023} (\ie the self-consistent $\Theta_e = \Theta_e(\sigma,
\beta)$ model). Hence, while approach \textit{(i)} has two tuneable
coefficients ($R_{\rm high}$ and $R_{\rm low}$), approach \textit{(ii)}
does not have tuneable coefficients and is fully determined by the
properties of the plasma via $\sigma$ and $\beta$.

Our comparison has been carried out over three different observational
aspects: the large-scale morphology of the M87 jet at 86 GHz, its width
along the direction of propagation, and, finally, the broadband spectral
energy distribution from M87. In this way, we have ascertained that
despite its ``rigidity'', the fully constrained microscopic model
\textit{(ii)} provides a match to the observations that is equally good
if not better than that obtained by the customary model
\textit{(i)}. More specifically, when considering the jet morphology, the
self-consistent $\Theta_e = \Theta_e(\sigma,\beta)$ model exhibits a
broader and more extended structure with pronounced limb brightening, in
better agreement with GMVA observations of M87*.  This behaviour reflects
the increased magnetic reconnection activity that is expected in
turbulent plasmas.

In addition, both models reproduce the observed jet width at $86\,{\rm
  GHz}$ for distances $r \gtrsim 0.3\,{\rm mas}$ from the launching
origin, although they both underestimate it on horizon scales. When
comparing the synthetic predictions with the observations, corresponding
reduced $\chi^2$ values are 1.46 for the phenomenological model and 1.39
for the self-consistent model. Finally, the spectral energy distribution
from radio to soft X-rays band (first synchrotron hump) also shows good
agreement between models and observations.  In particular, both models
match multi-epoch and recent simultaneous observations by EHT
collaboration, while the self-consistent model exhibits a deeper
self-absorption feature below $22\, {\rm GHz}$.

In summary, our results highlight the necessity and usefulness of
incorporating self-consistent models based on first-principles
simulations of collisionless plasma. While more complex, this approach
has two important advantages. First, it removes the presence of tuneable
parameters that can lead to degenerate imaging and hence to degenerate
theoretical interpretations. Second, in virtue of its ``rigidity'' it
makes the mapping from the observations to the physical conditions of the
plasma -- \ie the inverse problem of going from a radio image to a
distribution of $\beta$ and $\sigma$ -- simpler and less prone to
ambiguities. While we believe this is an important first step in the
direction of a more robust physical foundation for modelling BH shadows
and relativistic jets, a number of improvements are possible. In
particular, since all microscopic studies so far have been carried out in
flat spacetimes and with electron-proton mixtures only, it is clear that
general-relativistic PIC simulations that take into account also the
contribution of the spacetime curvature and the presence of multiple
charged species (\eg electrons, protons and positrons) will increase the
accuracy of the theoretical modelling and will lead to a more realistic
description.

%\begin{acknowledgments}
\section*{acknowledgments}
This research was supported by DGAPA-UNAM (grant IN110522) and the
Ciencia B\'asica y de Frontera 2023-2024 program of SECIHTI M\'exico
(projects CBF2023-2024-1102 and 257435); by the European Horizon Europe
Staff Exchange (SE) programme HORIZON-MSCA2021-SE-01 under Grant
No. NewFunFiCO-101086251; by the ERC Advanced Grant JETSET: Launching,
propagation and emission of relativistic jets from binary mergers and
across mass scales (Grant No. 884631); by the Deutsche
Forschungsgemeinschaft (DFG, German Research Foundation) through CRC-TR
211 Strong-interaction matter under extreme conditions (project number
315477589). C.M.F. is supported by the DFG research grant “Jet
physics on horizon scales and beyond" (Grant No. 443220636) within the DFG
research unit “Relativistic Jets in Active Galaxies" (FOR 5195).
S.S. acknowledges support from ``ICSC - Centro Nazionale di
Ricerca in High Performance Computing, Big Data and Quantum Computing'',
funded by European Union - NextGenerationEU.  
A.N. was supported by the Hellenic Foundation for Research and Innovation 
(ELIDEK) under Grant No 23698. L.R. is grateful to the
Theory Division at CERN, where part of this research was carried out and
to the Walter Greiner Gesellschaft zur F\"orderung der physikalischen
Grundlagenforschung e.V. through the Carl W. Fueck Laureatus Chair.
%\end{acknowledgments}

\newpage
\vspace{5mm}

%\bibliography{main} 

\end{document}